\documentclass[seceq]{ptptex}

\usepackage{graphicx}
\usepackage{wrapft}

\title{Effects of Impurities\\
in Quasi-One-Dimensional $S=1$ Antiferromagnets
}

\author{
Munehisa \textsc{Matsumoto}$^{1}$ and Hajime \textsc{Takayama}$^{2}$%
}

\inst{
$^1$Department of Physics, Tohoku University, Sendai 980-8578\\
$^2$Institute for Solid State Physics, University of Tokyo, Kashiwa 277-8581
}

\abst{
For the weakly coupled $S=1$ antiferromagnetic Heisenberg chains
on a simple cubic lattice, the effects of magnetic impurities
are investigated by the quantum Monte Carlo method with the
continuous-time loop algorithm. The transition temperatures of
the impurity-induced phase transitions for magnetic impurities
with $S=1/2$, $3/2$, and $2$ are determined and compared with
the transition temperature induced by the non-magnetic impurities.
Implications on the experimental results are discussed.
}

\begin{document}

\maketitle

\section{Impurity-Induced Phase Transitions}

There have been extensive studies
on the phase transitions caused
by magnetic fields and/or impurities
in the low-dimensional quantum gapped magnets,
which do not show phase transitions
down to zero temperature. Among them,
the {\it non-magnetic}-impurity-induced phase transitions
in the spin-Peierls material CuGeO$_3$\cite{hase}
and the Haldane material PbNi$_2$V$_2$O$_8$\cite{uchiyama}
have been experimentally investigated and the
temperature-dependence of the impurity-induced transition temperature has been
determined.

Recently in the experiment on PbNi$_2$V$_2$O$_8$,
the transition temperature induced
by {\it magnetic} impurities were
also reported and it was found that the impurity-induced
transition temperature shows strange non-monotonic dependence on
the magnitude of the impurity spin~\cite{imai}. Specifically,
the transition temperature caused by $S=1/2$ Cu$^{2+}$ ions
is very low compared with that by $S=0$ impurity Mg$^{2+}$ ions.
Furthermore, $S=3/2$ Co$^{2+}$ ions
induce by far the highest transition temperature
and the transition temperature induced by $S=5/2$ Mn$^{2+}$ impurities is
again lower than that induced by $S=0$ Mg$^{2+}$ impurities.
Motivated by these interesting experimental
results, we study the magnetic-impurity-induced phase
transition and determine the impurity-induced
transition temperature in
the quasi-one-dimensional $S=1$ antiferromagnetic
Heisenberg model by the quantum Monte Carlo simulations.
Based on the simulational results,
we discuss the picture for the
impurity-induced ordered state hoping to understand
the experimental results.

\section{Model}
\label{model}

As a simple model to describe the physics of static impurities
in the Haldane-gapped state, we take the weakly coupled
$S=1$ antiferromagnetic Heisenberg chains on a simple
cubic lattice with the host $S=1$ spins randomly replaced
by impurities with spin $S\ne 1$.
The Hamiltonian is written as follows.
\begin{equation}
{\cal H}=J\sum_{x,y,z}{\mathbf S}_{x,y,z}\cdot{\mathbf S}_{x+1,y,z}
+J'\left(\sum_{x,y,z}{\mathbf S}_{x,y,z}\cdot{\mathbf S}_{x,y+1,z}
+\sum_{x,y,z}{\mathbf S}_{x,y,z}\cdot{\mathbf S}_{x,y,z+1}\right)
\end{equation}
Here the strength of the intrachain coupling is $J$ and this parameter
is used as a unit to measure the energy and the temperature.
The strength of the interchain
coupling is $J'$ and we set the $x$-axis parallel to the coupled chains.

The model paramaters and simulational conditions are as follows.
It is known that the one-dimensional Haldane gap
at $J'=0$ survives even under the presence
of a three-dimensional interchain coupling
if its strength $J'$ is small enough\cite{sakai},
and we set $J'=0.01J$ with which the ground-state system is known to be
in the Haldane gapped phase\cite{sakai,koga}.
The critical value $J_{\rm c}'$ between the Haldane gapped phase
and the antiferromagnetic phase is estimated
to be $0.013$ by the mean-field theory
on the interchain coupling\cite{sakai}, which is expected
to give the lower bound of $J_{\rm c}'$,
and the value determined by the series-expansion method is
$J_{\rm c}'=0.026\pm 0.001$\cite{koga}. Practically it is not
sufficient just to satisfy $J'<J_{\rm c}'$, but we should also
take care to keep the gap of the pure system
small for the observation
of the impurity-induced phase transition at a
reasonably high temperature. That is why
we set $J'=0.01$ and not, say, $J'=0.001$.
We will explain this point in more detail in Sec.~\ref{discussions}.

For the system with $J'=0.01$,
we study the finite-temperature phase transition
induced by doped impurity spins
with $S=1/2$, $3/2$, and $2$,
randomly replacing the host spins that have $S=1$.
We utilize the quantum Monte Carlo method
with the continuous-time loop algorithm\cite{evertz}
and the subspin symmetrization technique\cite{todo}
for the efficient simulation of systems with general spin magnitude.
The sizes of the systems we simulated are $(L_x,L_y,L_z)=(8,8,8)$,
$(16,16,16)$, and $(32,32,32)$,
where the $L_x$, $L_y$, and $L_z$ are the number
of sites along the $x$, $y$, and $z$ axes, respectively.
For each system size and temperature,
we took average over 50 random samples.

In the present study, we fix the concentration of impurities
at 10\% and determine the magnetic-impurity-induced
transition N\'{e}el temperature and compare them with
the non-magnetic-impurity-induced N\'{e}el temperature.

\section{Results}
\label{results}

\begin{figure}[htb]
  \parbox{\halftext}{
    \centerline{\scalebox{0.5}{\includegraphics{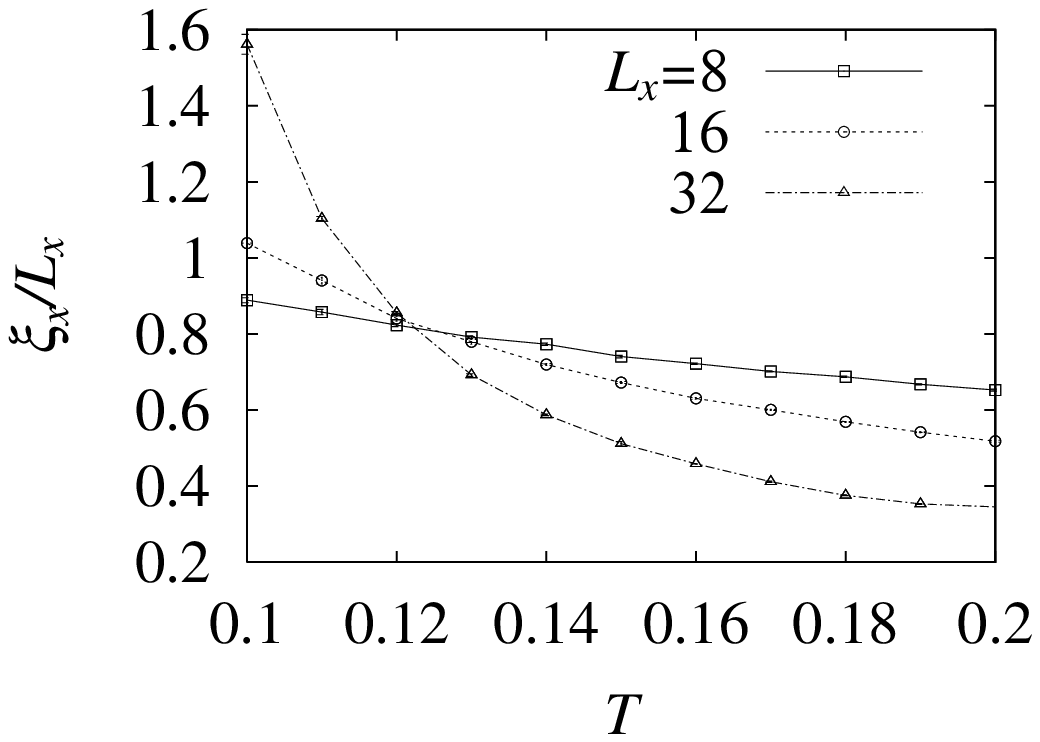}}}
    \caption{The determination of the transition temperature by the
      crossing point of the ratio of the correlation length $\xi_x$
      to the linear system size for system sizes $L_x=8$, $16$, and $32$,
      where $\xi_x$ and $L_x$ are the correlation length and the number of
      sites, respectively, along the $x$ axis.
      Here plotted are the data for the model doped with 10\% $S=2$ impurities.}
    \label{determination_TN}}
  \hfill
  \parbox{\halftext}{
    \centerline{\scalebox{0.5}{\includegraphics{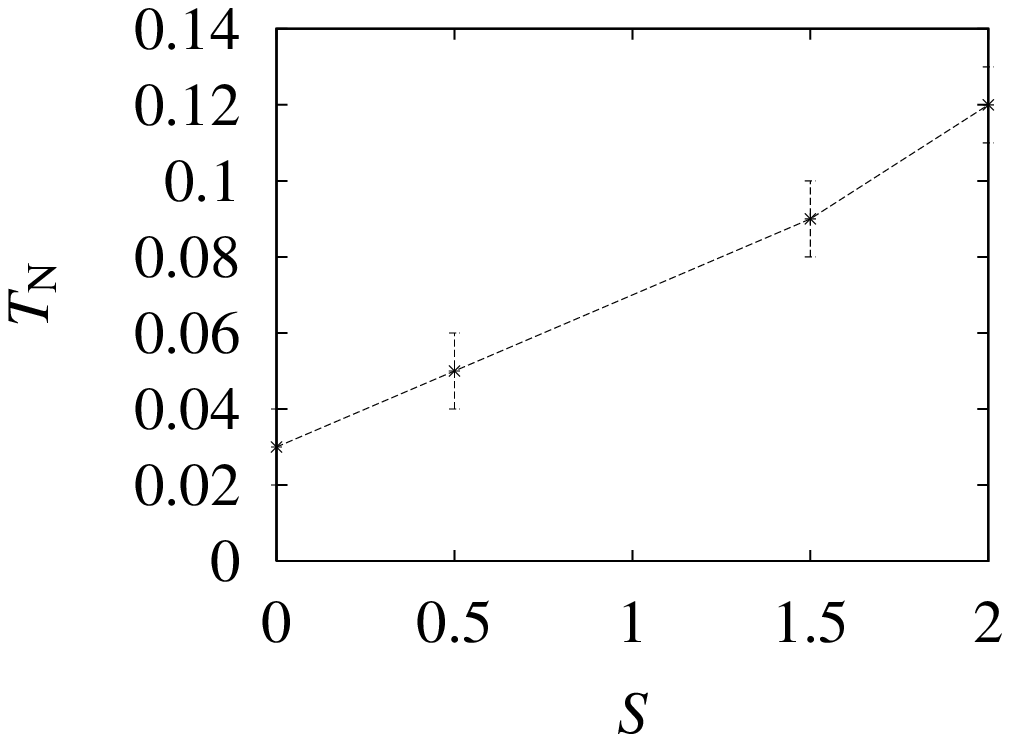}}}
    \caption{Impurity-induced temperature plotted against the magnitude of the
      impurity spin. The concentration of the impurities is fixed to be 10\%.}
    \label{TN_vs_S}}
\end{figure}

We show the way how we detemine the impurity-induced transition temperature
and its dependence on the spin magnitude of the impurities.
By the analogy with the standard finite-size scaling analysis,
we plot the ratio of the correlation length $\xi$ to the linear system size $L$
with respect to temperature varying the system size
and observe the crossing point in $\xi/L$ to determine
the transition temperature. Here the antiferromagnetic
correlation length is calculated
by the second moment method\cite{cooper} using
the dynamic staggered correlation function.
The analysis for the $S=2$ impurities is shown in Fig.~\ref{determination_TN}.
The impurity-induced N\'{e}el temperature
is determined to be $T_{\rm N}=0.12\pm 0.01$.
Thus determined impurity-induced transition temperatures are plotted
against the spin magnitude of the impurities in Fig.~\ref{TN_vs_S}.
This figure includes the results for the non-magnetic impurities.
As we can see, the $S=1/2$ impurities give higher transition temperature
than the non-magnetic impurities. Further, as we increase the spin magnitude
of the impurities, the transition temperatures increase monotonically
in contrast to the experimental results.

\section{Discussions I: Interpretation of the Simulational Results}
\label{discussions}

\begin{figure}
\centerline{\scalebox{0.5}{\includegraphics{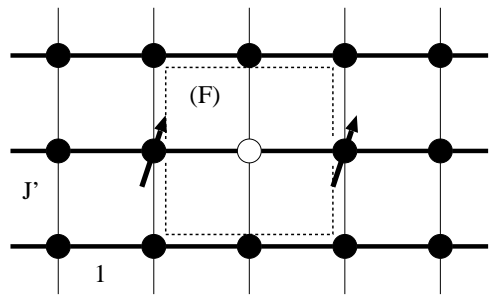}}}
\caption{Schematic figure of the site-dilution-induced
antiferromagnetic state. The superexchange interaction strength
along the chain is set to be unity and that along the interchain direction
is $J'$. In this figure, only the cross section on the $xy$ plane is shown.
The coupled chains are parallel to the $x$-axis.
The empty circle denotes the diluted site and
the filled one does the $S=1$ spins. The arrows denote the
effective $S=1/2$ spins that appear on
both of the nearest-neighboring sites next to the diluted site
along the chain. The interaction
between them is ferromagnetic along the shortest path that is indicated
as 'F' in the figure.}
\label{picture}
\end{figure}

To discuss the impurity-spin dependence of the impurity-induced
transition temperature for our simple model,
let us begin with the picture
for the non-magnetic-impurity-induced long-range ordered
phase\cite{sigrist,nagaosa,yasuda} which is
schematically shown in Fig.~\ref{picture}.
The ground state of the pure system is basically
understood as the valence-bond solid (VBS)\cite{aklt}-like state.
Non-magnetic impurities in this state act as cuts
in the Haldane chain
and the doped state can be regarded as a set of
finite-length $S=1$ Haldane chains with the open boundary condition.
On both edges of the finite length chain, there appear
effective spin degrees of freedom
with magnitude $S=1/2$\cite{aklt,miyashita}.
Thus on both neighboring sites of the diluted site,
$S=1/2$ effective spins appear and the interaction
between them along the shortest path is ferromagnetic
as shown in Fig.~\ref{picture}.
The long-range staggered correlation
between the effective spins is
supported by all of the interactions and thus
the non-magnetic impurities induce the
antiferromagnetic long-range order.

The typical energy of the interaction between the effective spins
decays exponentially with respect to the distance between
them with the decay constant proportional to the
gap of the pure system\cite{sigrist,nagaosa,yasuda}.
For the observation of the impurity-induced transition
at a reasonably high temperature, we need the gap of the
pure system to be sufficiently small to make large
the length scale of the correlation between the
effective spins that contribute to the long-range order.
So in the Haldane phase we choose a point near the phase boundary
as described in Sec.~\ref{model}, where the gap is almost
collapsing approaching the quantum critical point of the pure system.

This VBS-like picture reveals several subtle aspects
when we consider the effects of $S=1/2$ magnetic impurities.
At first glance, it might be expected that the $S=0$ impurities induce
higher transition temperature than $S=1/2$ impurities do
if we focus on the point that an effective $S=1/2$ spin
per $S=1/2$ impurity contribute to
the bulk long-range order while an effective $S=1$ spin per
$S=0$ impurity do. Here we imagine that the $S=1/2$ spin
in the position of the diluted site
in Fig.~\ref{picture} forms a singlet pair with one of
the effective spins. On the other hand, it is thought that
the opposite situation may well be realized
if we consider the paths of the interaction
of the effective spins that contribute to the bulk long-range order.
The $S=0$ spins cut the strong intrachain couplings and
only the extremely weak interchain couplings mediate the interaction
between the effective spins. Oppositely, the magnetic impurities keeps the
strong intrachain coupling by definition of the present model
and the strength of the interaction between the
local effective spins around the impurities is much larger.
Furthermore, the VBS structure might be broken around the
$S=1/2$ impurity spins and three $S=1/2$ spins per $S=1/2$ impurity spin
might contribute to the bulk antiferromagnetic order.
The simulational results showed that indeed the latter scenario
is more plausible for the impurity effects
in the present simple models. We should refine the naive VBS-like picture
by investigating the magnetic structure near the impurity sites.
This is the next problem now being studied.

\section{Discussions II: Toward the Understanding of the Experimental Results}
\label{discussions-2}

Obviously our model is too simple to describe the
real materials. One of the features we have not taken into account
is the anisotropy effects specific to each magnetic
impurity spins which might be playing a significant role in determining
the real impurity-induced transition temperature\cite{imai}. For example,
possible strong Ising-type anisotropy in $S=3/2$ Co$^{2+}$ impurities
could induce higher transition temperature than the others do\cite{imai}.
Another property we have missed is the next-nearest ferromagnetic coupling
that is known to be strong for PbNi$_2$V$_2$O$_8$\cite{imai,masuda}
and we should include this to give the more qualitatively precise results.
According to the discussions in the previous section,
the effects of next-nearest ferromagnetic coupling must be strong
as it enables the effective spins interact strongly even in the case of
$S=0$ impurities. Luckily we can do the quantum Monte Carlo simulations
without suffering from the sign problem\cite{troyer}
of the models with the next-nearest ferromagnetic coupling as
these additional interactions do not introduce frustration.
These are the next problems and now under investigation.

\section*{Acknowledgements}
One of the authors (M.) would like to thank Prof. S. Todo and Prof. T. Sakai
for valuable discussions
and Prof. K. Uchinokura and Prof. T. Masuda for useful comments.
The loop algorithm codes for the present calculations
are based on the library ``LOOPER version 2'' developed by
Prof. S. Todo and Dr. K. Kato and the codes for parallel simulations
are based on the library ``PARAPACK version 2'' developed by Prof. S. Todo.
The numerical calculations for the present work
were done on the SGI 2800 at the
Supercomputer center in the Institute for Solid
State Physics, University of Tokyo.

%

\end{document}